\documentclass[prl,twocolumn,10pt,showpacs]{revtex4-1}
\date{\today}

\newcommand{\insertplot}[5]{\begin{figure}
 \hfill\hbox to 0.05in{\vbox to #5in{\vfill
 \inputplot{#1}{#4}{#5}}\hfill}
 \hfill\vspace{-.1in}
 \caption{#2}\label{#3}
 \end{figure}}
 \newcommand{\inputplot}[3]{
 \special{ps: plotfile #1}
}
\newcounter{fig}   

\usepackage{epsfig}
\usepackage{amsmath}
\usepackage{amsfonts}
\usepackage{graphicx}
\usepackage[german, english]{babel}
\usepackage{amssymb}
\usepackage{ifthen}

\begin{document}

\newcommand{\dd}{\mbox{d}}
\newcommand{\tr}{\mbox{tr}}
\newcommand{\la}{\lambda}
\newcommand{\ta}{\theta}
\newcommand{\f}{\phi}
\newcommand{\vf}{\varphi}
\newcommand{\ka}{\kappa}
\newcommand{\al}{\alpha}
\newcommand{\ga}{\gamma}
\newcommand{\de}{\delta}
\newcommand{\si}{\sigma}
\newcommand{\bomega}{\mbox{\boldmath $\omega$}}
\newcommand{\bsi}{\mbox{\boldmath $\sigma$}}
\newcommand{\bchi}{\mbox{\boldmath $\chi$}}
\newcommand{\bal}{\mbox{\boldmath $\alpha$}}
\newcommand{\bpsi}{\mbox{\boldmath $\psi$}}
\newcommand{\brho}{\mbox{\boldmath $\varrho$}}
\newcommand{\beps}{\mbox{\boldmath $\varepsilon$}}
\newcommand{\bxi}{\mbox{\boldmath $\xi$}}
\newcommand{\bbeta}{\mbox{\boldmath $\beta$}}
\newcommand{\ee}{\end{equation}}
\newcommand{\eea}{\end{eqnarray}}
\newcommand{\be}{\begin{equation}}
\newcommand{\bea}{\begin{eqnarray}}
\newcommand{\ii}{\mbox{i}}
\newcommand{\e}{\mbox{e}}
\newcommand{\pa}{\partial}
\newcommand{\Om}{\Omega}
\newcommand{\vep}{\varepsilon}
\newcommand{\bfph}{{\bf \phi}}
\newcommand{\lm}{\lambda}
\def\theequation{\arabic{equation}}
\renewcommand{\thefootnote}{\fnsymbol{footnote}}
\newcommand{\re}[1]{(\ref{#1})}
\newcommand{\R}{{\rm I \hspace{-0.52ex} R}}
\newcommand{\N}{{\sf N\hspace*{-1.0ex}\rule{0.15ex}%
{1.3ex}\hspace*{1.0ex}}}
\newcommand{\Q}{{\sf Q\hspace*{-1.1ex}\rule{0.15ex}%
{1.5ex}\hspace*{1.1ex}}}
\newcommand{\C}{{\sf C\hspace*{-0.9ex}\rule{0.15ex}%
{1.3ex}\hspace*{0.9ex}}}
\newcommand{\eins}{1\hspace{-0.56ex}{\rm I}}
\renewcommand{\thefootnote}{\arabic{footnote}}

\title{Kink-antikink collisions in the $\phi^6$ model}

\author{Patrick Dorey$^{1,2}$, Kieran Mersh$^{1}$, 
Tomasz Romanczukiewicz$^{3}$ and Yasha Shnir$^{1}$}
\affiliation{$^{1}$Department of Mathematical Sciences, Durham
University, UK\\
$^{2}$LPTHE, CNRS -- Universit\'e Pierre et Marie Curie--Paris 6, 
4 Place Jussieu, Paris, France\\
$^{3}$Institute of Physics, Jagiellonian University, Krakow, Poland}

\begin{abstract}
We study kink-antikink
collisions in the one-dimensional non-integrable scalar $\phi^6$
model. Although the single-kink solutions for this model do not
possess an internal vibrational mode, our simulations reveal a
resonant scattering structure, thereby
providing a counterexample to the standard belief that existence 
of such a mode is a necessary
condition for multi-bounce resonances in general kink-antikink
collisions.  We investigate the two-bounce windows in detail,
and present evidence that this structure is caused by
existence of bound states in the spectrum of small oscillations about
a combined kink-antikink configuration.
\end{abstract}


\maketitle

%\medskip

%%%%%%%%%%%%%%%%%%%%%%%%%%%%%%%%%%%%%%%%%%%%%%%%%%%%%%%%%%%%%%%%%%
\noindent
{\it ~Introduction.~}
%%%%%%%%%%%%%%%%%%%%%%%%%%%%%%%%%%%%%%%%%%%%%%%%%%%%%%%%%%%%%%%
%
Solitary wave solutions in non-linear field theories have been studied
for many decades.
Perhaps the simplest examples are
the kink solutions which appear in models in one spatial dimension with a
potential with two or more degenerate minima. 
These models have 
applications in condensed matter physics \cite{Solitons},
field theory \cite{MantonSutcliffe,Vachaspati:2006zz} and as 
simplified models of domain wall collisions in cosmology
\cite{Vilenkin,Hawking:1982ga,%Anninos:1991un,
Giblin:2010bd}.
The double well case is exemplified by
the nonintegrable $\phi^4$ model.  
A particularly striking feature of this theory is the intricate structure
of kink-antikink scattering,
first observed numerically in the 1970s 
\cite{Ablowitz:1979aa}  
and much studied thereafter
\cite{Moshir:1981ja,Peyrard:1984qn,Campbell:1983xu,% Belova:1985fg,
Anninos:1991un,Goodman:2005,Goodman:2007}.
For initial velocities above a critical value $v_c\approx 0.2598$ 
the two incident
waves always escape to infinity after collision, with the emission 
of some radiation. Below $v_c$, the incident
waves generically become trapped, but 
there is also a complicated pattern of
narrow `resonance windows' within which
they are again able to escape to infinity.

The accepted explanation of the appearance of these windows,
both in this model and in similar theories such as the parametrically 
modified sine-Gordon model \cite{Peyrard:1984qn}, is that they
are related to a reversible exchange of energy 
between the translational and vibrational modes of the individual kinks
\cite{Campbell:1983xu}.
At the initial impact, some kinetic energy is transferred into
internal `shape' modes of the kink and antikink. They then separate
and propagate almost independently, but for initial velocities
less than $v_c$ they no longer have enough translational energy to
escape their mutual attraction, and so they return and collide a second
time. At this point some of the energy stored in
the shape modes can be returned to the translational modes, tipping the
energy balance back again and allowing the kink and antikink to escape
to infinity, provided that there is an appropriate resonance between
the interval between the two collisions, and the period of
the internal modes. More generally, sufficient energy might be
returned to the translational modes after three or more kink-antikink
collisions, leading to an intricate nested structure of resonance
windows, a picture which has been confirmed by both numerical and
analytical studies~\cite{Campbell:1983xu,Peyrard:1984qn,%
Anninos:1991un,Goodman:2005,Goodman:2007}.

For this mechanism to work, the kink and antikink must each support at
least one internal vibrational mode, within which energy can be 
stored prior to being transferred back to the translational modes 
after a number of bounces. This has led to the commonly-held belief,
stated for example in \cite{Goodman:2005},
that the existence of an internal kink mode is a {\em
necessary}\/ condition for the appearance of resonance windows. The
example of the parametrically modified sine-Gordon model lends further
support to this view: depending on the value of a parameter, kinks 
and antikinks do or do not possess an internal mode; correspondingly,
resonance windows do or do not appear~\cite{Peyrard:1984qn}.
Resonances have also been observed in vector soliton
collisions \cite{Yang:2000,Goodman:2005b} and in the scattering of
kinks on impurities~\cite{Kivshar:1991zz}, but again,
the mechanism always relies on the presence of a localised internal mode, 
either of a single kink or of an impurity, or both.

In the present Letter we revisit this question in the
context of kink-antikink scattering in the $\phi^6$ model
\cite{Lohe:1979mh}. The potentials mentioned in
\cite{Peyrard:1984qn} include a general polynomial form, but 
the $\phi^6$ model has a number of features whose relevance to
resonant scattering have not been appreciated to date,
even in the recent work \cite{Hoseinmardy:2010zz}.
As for the $\phi^4$
model, it is non-integrable, but, as for the
modified sine-Gordon model in the regime of no resonant scattering, 
a single $\phi^6$ kink does not have an internal oscillatory mode
\cite{Lohe:1979mh}. Nevertheless, certain
kink-antikink collisions
exhibit resonant scattering, thus providing
a counterexample to the standard lore.
We elucidate the mechanism for this process, finding that
the underlying reason is a reversible
transfer of kinetic energy from the kinks to a collective bound
state trapped by the $\bar KK$ pair.  

\smallskip

%%%%%%%%%%%%%%%%%%%%%%%%%%%%%%%%%%%%%%%%%%%%%%%%%%%%%%%%%%%%%%%%%%
\noindent
{\it ~The model.~}
%%%%%%%%%%%%%%%%%%%%%%%%%%%%%%%%%%%%%%%%%%%%%%%%%%%%%%%%%%%%%%%%%%
The one-dimensional $\phi^6$ theory can be defined by the
rescaled Lagrangian density \cite{Lohe:1979mh}
\be  
\label{Lag}
{\cal L} = \frac{1}{2}\partial_\mu \phi \partial^\mu \phi - 
\frac{1}{2}\phi^2\left(\phi^2 -1\right)^2.
\ee
The model has three vacua $\phi_v \in \{-1,0,1\}$. Static kinks and
antikinks, interpolating between neighbouring vacua,
can be found from the
one-kink solution
$\phi_K(x) \equiv \phi_{(0,1)}(x)= \sqrt{(1{+}\tanh x)/2}$
using the discrete symmetries 
of the model under
$\phi \to -\phi$ and/or $x\to -x$,
so $\phi_{\bar K}(x)\equiv\phi_{(1,0)}(x)= \phi_{K}(-x)$,
$\phi_{(0,-1)}(x)=-\phi_{K}(x)$ and $\phi_{(-1,0)}(x)=-\phi_{K}(-x)$.
These all have mass $M=1/4$.

The model also possesses perturbative meson states, 
fluctuations about the vacua $\phi=0,\,\pm 1$.
Fluctuations around the static solutions 
$\phi_s(x)$ which interpolate between these vacua can be
treated by setting $\phi(x,t) =
\phi_s(x) + \eta(x) e^{i\omega t}$. The
linearised field equations are $-\eta_{xx} + U(x) \eta = 
\omega^2 \eta$ where %\cite{Lohe:1979mh}.
the potential $U(x)$ is
\be \label{Pot}
U(x) = 15 \phi_s^4 - 12 \phi_s^2 +1.
\ee
For an isolated kink there
are no localised solutions to this equation
beyond the usual translational zero mode,
reflecting the absence of internal oscillatory
modes. The states
of the continuum spectrum can be written in terms of 
hypergeometric functions~\cite{Lohe:1979mh}.

\smallskip

%%%%%%%%%%%%%%%%%%%%%%%%%%%%%%%%%%%%%%%%%%%%%%%%%%%%%%%%%%%%%%%%%%
\noindent
{\it ~Numerical results.~}
%%%%%%%%%%%%%%%%%%%%%%%%%%%%%%%%%%%%%%%%%%%%%%%%%%%%%%%%%%%%%%%%%%
The initial conditions we took
correspond to a widely separated kink-antikink
pair propagating towards a collision point.
To find a numerical solution of the PDE describing the evolution of
the system, we used a pseudo-spectral method on a grid
containing 2048 nodes with periodic boundary conditions.  For the time
stepping function we used a symplectic (or geometric) integrator of
8$^{\rm th}$
order to ensure that the energy is conserved.  The time and the
spatial steps were $\delta x = 0.25$ and $\delta t = 0.025$.
To check numerical stability we repeated selected calculations 
with $\delta t=0.05$ and $\delta t=0.0125$.

Since the model \re{Lag} contains two distinct classes of 
vacua, $0$ and $\pm 1$, kink-antikink
collisions in the sectors built on the
$\phi_v=0$ and the $\phi_v=\pm 1$ vacua have to be analysed
separately.
In the former case the initial configuration,
which we denote as $K\bar K$ or
$(0,1)+(1,0)$, can be taken as a superposition
$\phi(x)=\phi_K(x{+}a) + \phi_{\bar K}(x{-}a)-1$ where $a>0$ is the
initial (half-)separation parameter, 
which we set equal to $12.5$; in the latter case we take
$\phi(x)=\phi_{\bar K}(x{+}a) + \phi_{K}(x{-}a)$ and
denote this as $\bar KK$ or $(1,0)+(0,1)$.
In neither case are there internal vibrational
modes bound to the individual kinks. By the standard 
picture, we would therefore
predict that neither should exhibit any resonance structure.

\begin{figure}
 \begin{center}
 \includegraphics[width=9cm,angle=0,bb=50 50 554 698]{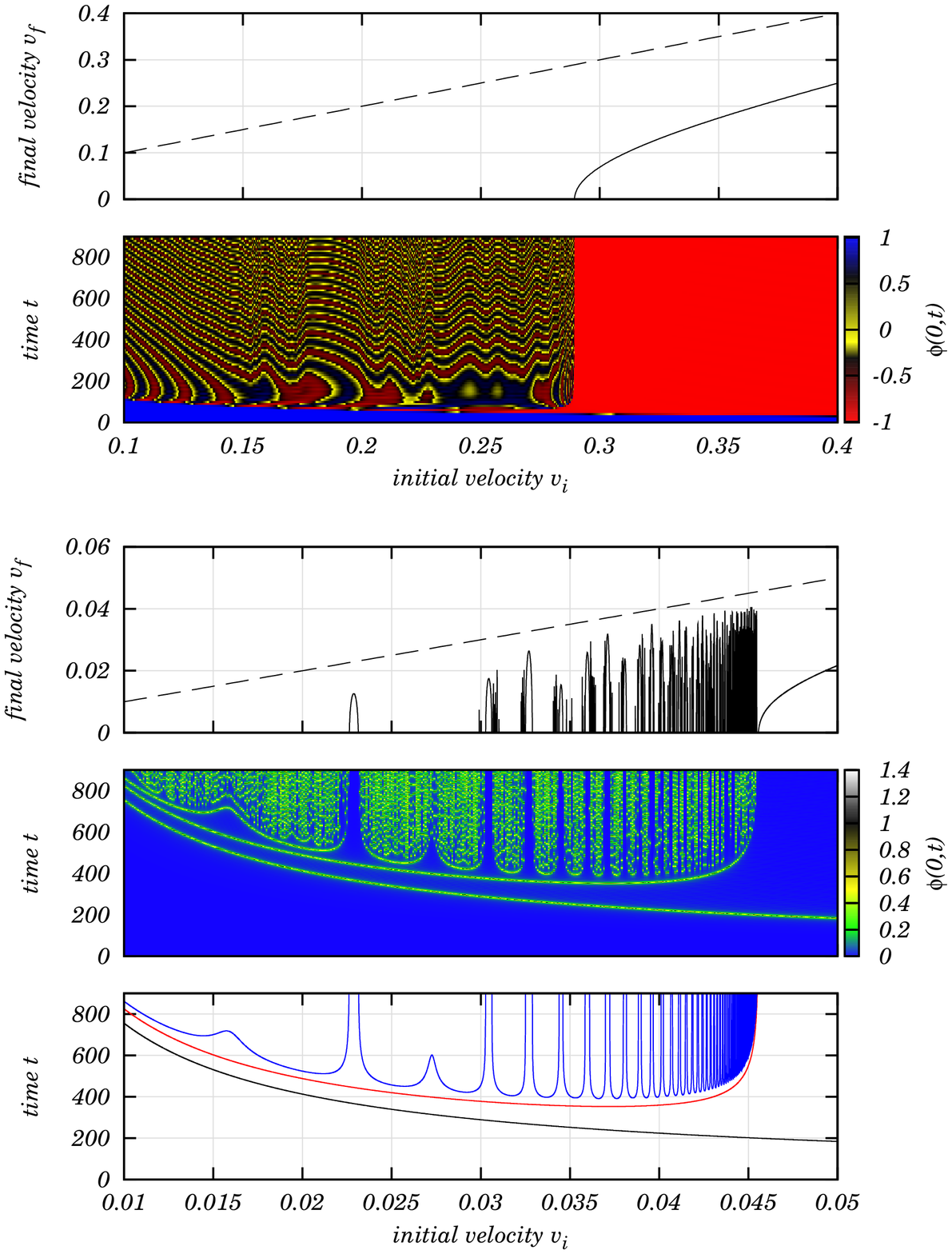}
 \caption{\label{fig:1}
\rlap{\vbox to 0pt{\vskip -298pt\hskip -37pt \footnotesize (a)}}%
\rlap{\vbox to 0pt{\vskip -244pt\hskip -37pt \footnotesize (b)}}%
\rlap{\vbox to 0pt{\vskip -158pt\hskip -37pt \footnotesize (c)}}%
\rlap{\vbox to 0pt{\vskip -103pt\hskip -37pt \footnotesize (d)}}%
\rlap{\vbox to 0pt{\vskip -40pt\hskip -37pt \footnotesize (e)}}%
 \small $\{K,\bar K\}$ collisions in the 
$(0,1)+(1,0)$ ((a), (b)) and $(1,0)+(0,1)$ sectors
((c), (d), (e)). 
Plots (a) and (c) show the fitted final velocity of the kink
as a function of initial velocity, with the dotted line indicating
the result for a purely elastic collision, while (b) and (d) depict
the field values measured at the collision centre. The final plot,
(e), shows the times to the first, second and third kink-antikink 
collisions in the $(1,0)+(0,1)$ sector.\\[-22pt]
}
\end{center}
\end{figure}

In the case of $(0,1)+(1,0)$ $K\bar K$ collisions, this is indeed what 
we found (Figs.~\ref{fig:1}(a), (b)).
For $v < v_{cp} \approx 0.289$ the pair always become trapped,
while for $v > v_{cp}$ the collision yields a reflected pair
of solitons together with some radiation: $(0,1)+(1,0) \to
(0,-1)+(-1,0)$. However the $(1,0)+(0,1)$ $\bar K K$
collision reveals a very different picture, illustrated in
Figs.~\ref{fig:1}(c), (d) and (e).
An intricate pattern of escape
windows can be seen, up to a critical velocity
$v_{cr}\approx 0.0457$, after which the kinks always have enough energy to
separate.  This is very similar to
the behaviour of the $\phi^4$ model, even though
there are no internal
modes of the scattering kinks into which kinetic energy can be 
transferred. Thus, we have
to look for another explanation of our results.  

\smallskip

%%%%%%%%%%%%%%%%%%%%%%%%%%%%%%%%%%%%%%%%%%%%%%%%%%%%%%%%%%%%%%%%%%
{\it ~The mechanism.~}
%%%%%%%%%%%%%%%%%%%%%%%%%%%%%%%%%%%%%%%%%%%%%%%%%%%%%%%%%%%%%%%%%%
The spectrum of linear perturbations around a
single
$\phi^6$ kink differs from that for
the $\phi^4$ kink in that the potential $U(x)$
is not symmetrical with respect to reflections $x \to -x$. 
As a result the potential for a 
kink-antikink pair depends on the order in which they appear on the
line: a well-separated $K\bar K$ pair (Fig.\,\ref{fig:2}(a)) has a 
raised central plateau, while an equally separated $\bar KK$ pair
(Fig.\,\ref{fig:2}(b)) has instead a wide central well.
If the velocities of the kinks are relatively small, we expect -- and
will justify later -- that the adiabatic 
approximation can be used to find the spectrum of small fluctuations
about these two configurations, using
Eq.\,\re{Pot} for the appropriate choices of $\phi_s$.
Fig.\,\ref{fig:3}(b) shows our numerical 
results for the $(1,0)+(0,1)$ $\bar KK$ pair
as a function of separation parameter $a$. 
The two lowest states are quasizero modes, rapidly approaching zero
from opposite sides as the separation grows. On top of these is a
tower of meson states with separation-dependent energies. The 
analogous plot for the $(0,1)+(1,0)$ $K\bar K$ pair, by contrast,
shows only the two quasizero modes.

So, a new feature of $\bar KK$ collisions in the $\phi^6$ model
is that although there are no internal modes of the individual
kinks, energy can be stored after an initial impact
in the trapped meson states of the composite
$(1,0)+(0,1)$ configuration. This opens the possibility that, for
collisions with $v_i<v_{cr}$ satisfying a suitable resonance condition, 
this energy might be returned to the translational modes on a 
subsequent recollision, thereby allowing the kink and antikink to 
return to infinity.

\begin{figure}
   \centering
   \vskip -32pt
   \includegraphics[width=7.2cm,angle=0,bb=50 50 554 482]{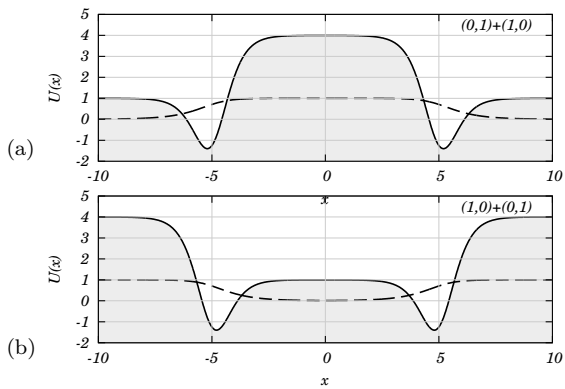} 
   \vskip -8pt
   \caption{\label{fig:2}
\rlap{\vbox to 0pt{\vskip -104pt\hskip -37pt \footnotesize (a)}}%
\rlap{\vbox to 0pt{\vskip -29pt\hskip -37pt \footnotesize (b)}}%
\small The potential $U(x)$ (solid lines) for
linear perturbations about the composite
   $(0,1)+(1,0)$ $K\bar K$ (a) and $(1,0)+(0,1)$ $\bar K K$
(b) configurations (dashed lines).\\[-12pt]
}
\end{figure}

\begin{figure}
 \centering
 \includegraphics[height=7.8cm,angle=270,bb=66 83 551 748]{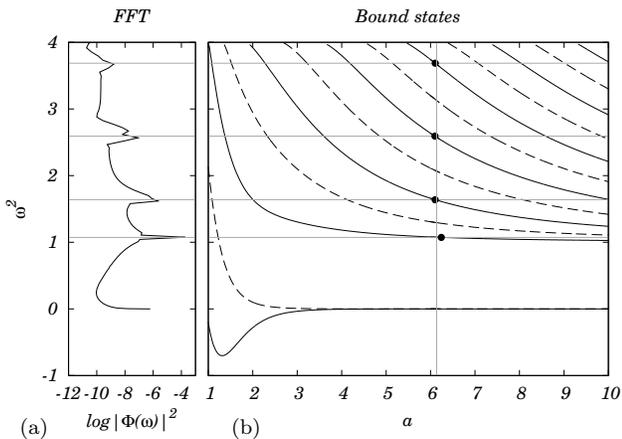}
   \vskip -2pt
\caption{\label{fig:3}
\rlap{\vbox to 0pt{\vskip -24pt\hskip -33pt \footnotesize (a)}}%
\rlap{\vbox to 0pt{\vskip -24pt\hskip 47pt \footnotesize (b)}}%
\small 
(a) Peaks of the Fourier transform of the field at the origin in the
two-bounce window for $v=0.04548$, vs.\
(b) even (solid) and odd (dashed) bound
states for the $(1,0)+(0,1)$ configuration, as a
function of the half-separation $a$.\\[-17pt]
}
\end{figure}

To support this picture, we first assign a `bounce number' to 
each window, equal to the number of collisions that the kink and
antikink undergo before their final escape to infinity. Within each
window the bounce number is constant.
The first two-bounce windows, discernible on
Fig.\,\ref{fig:1}(e) as the intervals of velocity
within which the time to the third collision diverges, are 
centred at $v_i\approx 0.0228$, followed by a `false' window at
$v_i\approx 0.0273$, then a third at $v_i\approx 0.0303$, and so on.
The regions near to these windows exhibit nested structures of 
higher-bounce windows, similar to those seen in kink-antikink 
interactions in the $\phi^4$ model~\cite{Anninos:1991un,Goodman:2005}. 

\begin{figure}
   \centering
   \vskip -2pt
   \includegraphics[height=8cm,angle=270]{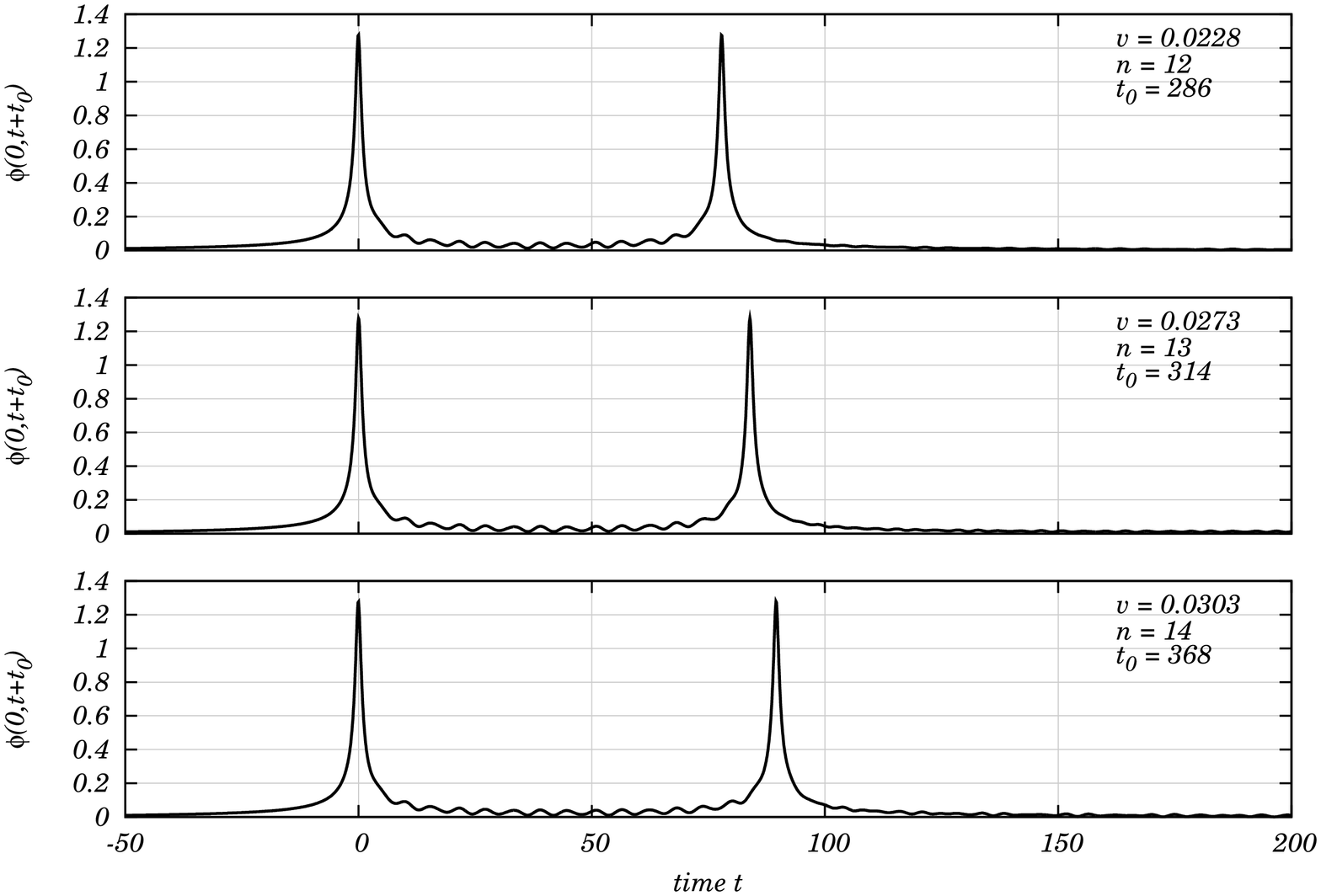}
   \vskip -2pt
   \includegraphics[height=7.95cm,angle=270]{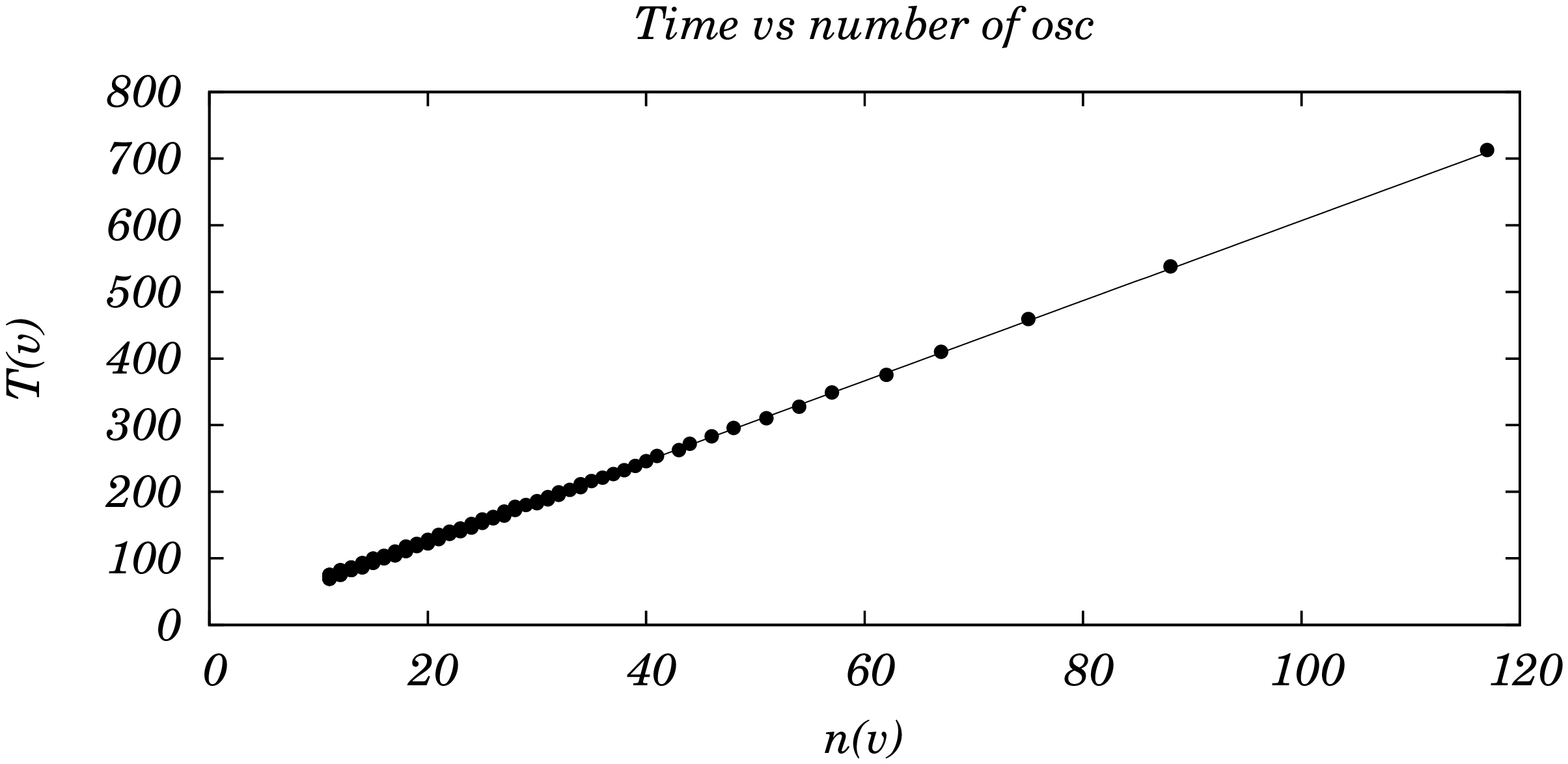}
   \vskip -1pt
 \caption{\label{fig:4}
\rlap{\vbox to 0pt{\vskip -146pt\hskip -37pt \footnotesize (a)}}%
\rlap{\vbox to 0pt{\vskip -42pt\hskip -37pt \footnotesize (b)}}%
\small (a) Field value at the collision centre as a function of time
from the first collision
for three consecutive two-bounce windows.
(b) The time between the two collisions in the
two-bounce windows vs.\
the number of oscillations.\\[-15pt]
}
\end{figure}

Our analysis implies that each two-bounce window should be associated
with a distinct integer $n$, the number of collective mode
oscillations between its two collisions.  In Fig.~\ref{fig:4}(a) we
illustrate this for the three windows just discussed. It is easily
checked that the number of oscillations of $\phi(0,t)$ between the two
collisions increases by one from one plot to the next.
Furthermore, our numerical results clearly indicate that these
oscillations are those of the lowest collective mode of the
$\bar KK$ system. In the two-bounce
windows we considered, we observed maximal kink-antikink separations 
between bounces up to $R\equiv 2a \approx 12$. 
Fig.~\ref{fig:3}(a) shows the Fourier transform of the value of
the field at the collision centre between the times of the
first two bounces
for $v_i=0.04548$. 
The time between these
bounces is $T \approx 660$, during most of which
the separation between the kinks is close to $12$.
The Fourier transform shows four peaks
at frequencies $1.045, 1.28, 1.61, 1.92$, with the strength
of the second peak about one order smaller than that of
the first, and the 
higher ones strongly suppressed.  The positions of these peaks 
relative to the energies of the first 4 parity-even modes 
(Fig.~\ref{fig:3}(a); note that the symmetry of the initial condition
rules out any coupling to parity-odd modes) are consistent with
a half-separation between the kink and antikink
of $a \approx 6.13$, in good agreement with the observed
value. This also lends support to our use of the adiabatic approximation in
the analysis of fluctuations about the kink-antikink configuration.

In Fig.~\ref{fig:4}(b) we show
the dependence of the times $T(v)$ between the two
collisions in the two-bounce windows on the
number $n(v)$ of oscillations in between these collisions.
The results are a good fit to the linear relation 
$\omega T= 2\pi n + \delta$, where 
$\omega = 1.0452$, the same as the frequency of the lowest collective mode 
at $2a\approx 12$ and the value
found from the Fourier transform at $v_i=0.04548$, and
$\delta= 6.0277$. Requiring $\delta$ to lie between $0$ and $2\pi$, as
here, fixes the number $n$ of collective mode oscillations assigned to
the three windows shown in Fig.~\ref{fig:4}(a) to be $12$,
$13$ and $14$ respectively, while the initial velocity
$v_i=0.04548$ used for the Fourier transform plot is very close to the 
$n=109$ two-bounce window.

A linear fit with this value of $\omega$ is exactly as predicted by a
resonance mechanism driven by the first mode, and closely resembles
similar relations in the $\phi^4$ model \cite{Campbell:1983xu},
although the detailed mechanism here is different.  In particular, in
resonant $\phi^4$ scattering, energy is exponentially localised on the
kink and antikink, while
in the $\phi^6$ theory this energy is stored in an extended
meson state residing in the potential well formed between the kink 
and antikink. This results in a small amount of
radiation pressure on the kinks after their initial impact,
which we observed for $v_i=0.05$, just larger than $v_c$, in a 
small (of order $10^{-6}$) 
acceleration of the bouncing kinks away from each other, when measured
at times $t\approx 300$ after collision so that the kinks had become
well separated. A corresponding simulation in $\phi^4$ theory 
showed that any such effect is at least $10^4$ times weaker there.

Nevertheless, we can pursue the similarity 
between the two models by considering the asymptotic 
attractive forces between kink and antikink in the absence of radiation
pressure, which  are $F(a)
\sim 2 e^{-R}$ for the $\bar K K$ pair, and $F(a) \sim 2 e^{-2R}$
for $K\bar K$ pair, where $R=2a$ is the kink-antikink separation. 
The arguments of \cite{Campbell:1983xu} would then predict
$T(v) \propto (v_{cr}^2-v_i^2)^{-\alpha}$ for impact velocities $v_i$ just
below $v_{cr}$, with $\alpha=0.5$. Our fits based on this functional
form favour a smaller value for $\alpha$, in the range $0.39 - 0.45$.
It would be interesting to see whether a more refined treatment,
taking radiation pressure into account, could account for these
discrepancies.

Perhaps the most intriguing feature of resonant $\phi^6$ scattering,
though, is the `missing' window at $n=13$. For resonant $\phi^4$
scattering, two-bounce windows 
are also missing, for $n<3$, but once they set in
they are found for all $n$, at least up to initial velocities very
close to $v_c$ \cite{Campbell:1983xu}. By contrast, in $\phi^6$
scattering we found the first two-bounce window at $n=12$, then a false window
at $n=13$, and then true windows for all higher values of $n$ that we
examined. Even more remarkably, a similar structure is
reproduced when looking at the three-bounce windows next to a given
two-bounce window, and we suspect that this pattern will continue at
all higher levels. It is possible that an explanation for this
behaviour will be found in a careful treatment of the higher modes of
the bound-state spectrum, but it remains a major challenge to convert
this idea into a robust set of predictions.

\smallskip

%%%%%%%%%%%%%%%%%%%%%%%%%%%%%%%%%%%%%%%%%%%%%%%%%%%%%%%%%%%%%%%%%%
\noindent
{\it ~Conclusions.~}
The intricate structure of $\phi^6$ kink collisions
overturns some previous beliefs
about resonant kink scattering, and shows that it has wider
relevance than previously thought. The new mechanism enabling
resonances to occur is the formation of meson bound states in
the potential well created in the space {\em between}\/ the
constituents of a 
suitably-ordered kink-antikink pair.
The resulting windows are less regular than for $\phi^4$ scattering, 
and exhibit gaps.  For large kink-antikink separations
we also saw an additional long-range interaction through
the trapped meson field, visible as radiation pressure. 

\smallskip

We thank M.\ Peyrard for helpful discussions, and Hadi Susanto and Roy
Goodman for bringing references \cite{Hoseinmardy:2010zz} and
\cite{Yang:2000,Goodman:2005b} respectively to our attention after
the first version of this Letter had been placed on the electronic
archives.
PED thanks LPTHE and YS 
thanks the University of Oldenburg for hospitality.  The work was 
supported in part
by the STFC (PED, KM), the CNRS (PED), and the A.\ von Humboldt
Foundation (YS).
%%%%%%%%%%%%%%%%%%%%%%%%%%%%%%%%%%%%%%%%%%%%%%%%%%%%%%%%%%%%%%%%%%
\begin{small}

\end{small}

\end{document}